\documentclass[a4paper,11pt]{article}
\pdfoutput=1 
\usepackage[T1]{fontenc} 

\makeatletter
\gdef\@fpheader{ }
\gdef\@journal{ }
\makeatother
\RequirePackage{amsmath}
\RequirePackage{amssymb}
\RequirePackage{epsfig}
\RequirePackage{graphicx}
\RequirePackage[numbers,sort&compress]{natbib}
\RequirePackage{color}
\RequirePackage[colorlinks=true
,urlcolor=blue
,anchorcolor=blue
,citecolor=blue
,filecolor=blue
,linkcolor=blue
,menucolor=blue
,pagecolor=blue
,linktocpage=true
,pdfproducer=medialab
,pdfa=true
]{hyperref}

\newif\ifnotoc\notocfalse
\newif\ifemailadd\emailaddfalse
\newif\iftoccontinuous\toccontinuousfalse
\makeatletter
\def\@subheader{\@empty}
\def\@keywords{\@empty}
\def\@abstract{\@empty}
\def\@xtum{\@empty}
\def\@dedicated{\@empty}
\def\@arxivnumber{\@empty}
\def\@collaboration{\@empty}
\def\@collaborationImg{\@empty}
\def\@proceeding{\@empty}
\def\@preprint{\@empty}

\newcommand{\subheader}[1]{\gdef\@subheader{#1}}
\newcommand{\keywords}[1]{\if!\@keywords!\gdef\@keywords{#1}\else%
\PackageWarningNoLine{\jname}{Keywords already defined.\MessageBreak Ignoring last definition.}\fi}
\renewcommand{\abstract}[1]{\gdef\@abstract{#1}}
\newcommand{\dedicated}[1]{\gdef\@dedicated{#1}}
\newcommand{\arxivnumber}[1]{\gdef\@arxivnumber{#1}}
\newcommand{\proceeding}[1]{\gdef\@proceeding{#1}}
\newcommand{\xtumfont}[1]{\textsc{#1}}
\newcommand{\correctionref}[3]{\gdef\@xtum{\xtumfont{#1} \href{#2}{#3}}}
\newcommand\jname{JHEP}
\newcommand\acknowledgments{\section*{Acknowledgments}}

\newcommand\preprint[1]{\gdef\@preprint{\hfill #1}}

\makeatother

\newcommand\note[2][]{%
\if!#1!%
\stepcounter{footnote}\footnotetext{#2}%
\else%
{\renewcommand\thefootnote{#1}%
\footnotetext{#2}}%
\fi}

\makeatletter
\newtoks\auth@toks
\renewcommand{\author}[2][]{%
  \if!#1!%
    \auth@toks=\expandafter{\the\auth@toks#2\ }%
  \else
    \auth@toks=\expandafter{\the\auth@toks#2$^{#1}$\ }%
  \fi
}
\makeatother
\makeatletter
\newtoks\affil@toks\newif\ifaffil\affilfalse
\newcommand{\affiliation}[2][]{%
\affiltrue
  \if!#1!%
    \affil@toks=\expandafter{\the\affil@toks{\item[]#2}}%
  \else
    \affil@toks=\expandafter{\the\affil@toks{\item[$^{#1}$]#2}}%
  \fi
}
\makeatother
\makeatletter
\newtoks\email@toks\newcounter{email@counter}%
\newcommand{\emailAdd}[1]{%
\emailaddtrue%
\ifnum\theemail@counter>0\email@toks=\expandafter{\the\email@toks, \@email{#1}}%
\else\email@toks=\expandafter{\the\email@toks\@email{#1}}%
\fi\stepcounter{email@counter}}
\newcommand{\@email}[1]{\href{mailto:#1}{\tt #1}}
\makeatother

\makeatletter
\newcommand*\collaboration[1]{\gdef\@collaboration{#1}}
\newcommand*\collaborationImg[2][]{\gdef\@collaborationImg{#2}}
\makeatletter
\newcommand\afterLogoSpace{\smallskip}
\newcommand\afterSubheaderSpace{\vskip3pt plus 2pt minus 1pt}
\newcommand\afterProceedingsSpace{\vskip21pt plus0.4fil minus15pt}
\newcommand\afterTitleSpace{\vskip23pt plus0.06fil minus13pt}
\newcommand\afterRuleSpace{\vskip23pt plus0.06fil minus13pt}
\newcommand\afterCollaborationSpace{\vskip3pt plus 2pt minus 1pt}
\newcommand\afterCollaborationImgSpace{\vskip3pt plus 2pt minus 1pt}
\newcommand\afterAuthorSpace{\vskip5pt plus4pt minus4pt}
\newcommand\afterAffiliationSpace{\vskip3pt plus3pt}
\newcommand\afterEmailSpace{\vskip16pt plus9pt minus10pt\filbreak}
\newcommand\afterXtumSpace{\par\bigskip}
\newcommand\afterAbstractSpace{\vskip16pt plus9pt minus13pt}
\newcommand\afterKeywordsSpace{\vskip16pt plus9pt minus13pt}
\newcommand\afterArxivSpace{\vskip3pt plus0.01fil minus10pt}
\newcommand\afterDedicatedSpace{\vskip0pt plus0.01fil}
\newcommand\afterTocSpace{\bigskip\medskip}
\newcommand\afterTocRuleSpace{\bigskip\bigskip}
\newlength{\affiliationsSep}
\newcommand\beforetochook{\pagestyle{myplain}\pagenumbering{roman}}

\DeclareFixedFont\trfont{OT1}{phv}{b}{sc}{11}

\renewcommand\maketitle{
\pagestyle{empty}
\thispagestyle{titlepage}
\setcounter{page}{0}
\noindent{\small\scshape\@fpheader}\@preprint\par

\afterLogoSpace
\if!\@subheader!\else\noindent{\trfont{\@subheader}}\fi
\afterSubheaderSpace
\if!\@proceeding!\else\noindent{\sc\@proceeding}\fi
\afterProceedingsSpace
{\LARGE\flushleft\sffamily\bfseries\@title\par}
\afterTitleSpace
\hrule height 1.5\p@%
\afterRuleSpace
\if!\@collaboration!\else
{\Large\bfseries\sffamily\raggedright\@collaboration}\par
\afterCollaborationSpace
\fi
\if!\@collaborationImg!\else
{\normalsize\bfseries\sffamily\raggedright\@collaborationImg}\par
\afterCollaborationImgSpace
\fi
{\bfseries\raggedright\sffamily\the\auth@toks\par}
\afterAuthorSpace
\ifaffil\begin{list}{}{%
\setlength{\leftmargin}{0.28cm}%
\setlength{\labelsep}{0pt}%
\setlength{\itemsep}{\affiliationsSep}%
\setlength{\topsep}{-\parskip}}
\itshape\small%
\the\affil@toks
\end{list}\fi
\afterAffiliationSpace
\ifemailadd 
\noindent\hspace{0.28cm}\begin{minipage}[l]{.9\textwidth}
\begin{flushleft}
\textit{E-mail:} \the\email@toks
\end{flushleft}
\end{minipage}
\else 
\PackageWarningNoLine{\jname}{E-mails are missing.\MessageBreak Plese use \protect\emailAdd\space macro to provide e-mails.}
\fi
\afterEmailSpace
\if!\@xtum!\else\noindent{\@xtum}\afterXtumSpace\fi
\if!\@abstract!\else\noindent{\renewcommand\baselinestretch{.9}\textsc{Abstract:}}\ \@abstract\afterAbstractSpace\fi
\if!\@keywords!\else\noindent{\textsc{Keywords:}} \@keywords\afterKeywordsSpace\fi
\if!\@arxivnumber!\else\noindent{\textsc{ArXiv ePrint:}} \href{http://arxiv.org/abs/\@arxivnumber}{\@arxivnumber}\afterArxivSpace\fi
\if!\@dedicated!\else\vbox{\small\it\raggedleft\@dedicated}\afterDedicatedSpace\fi
\ifnotoc\else
\iftoccontinuous\else\newpage\fi
\beforetochook\hrule
\tableofcontents
\afterTocSpace
\hrule
\afterTocRuleSpace
\fi
\setcounter{footnote}{0}
\pagestyle{myplain}\pagenumbering{arabic}
} 

\renewcommand{\baselinestretch}{1.1}\normalsize
\divide\@tempdima\baselineskip
\@tempcnta=\@tempdima
\skip\@mpfootins = \skip\footins
\renewcommand{\@dotsep}{10000}

\newcommand\ps@myplain{
\pagenumbering{arabic}
\renewcommand\@oddfoot{\hfill-- \thepage\ --\hfill}
\renewcommand\@oddhead{}}
\let\ps@plain=\ps@myplain

\newcommand\ps@titlepage{\renewcommand\@oddfoot{}\renewcommand\@oddhead{}}

\numberwithin{equation}{section}

\renewcommand\section{\@startsection{section}{1}{\z@}%
                                   {-3.5ex \@plus -1.3ex \@minus -.7ex}%
                                   {2.3ex \@plus.4ex \@minus .4ex}%
                                   {\normalfont\large\bfseries}}
\renewcommand\subsection{\@startsection{subsection}{2}{\z@}%
                                   {-2.3ex\@plus -1ex \@minus -.5ex}%
                                   {1.2ex \@plus .3ex \@minus .3ex}%
                                   {\normalfont\normalsize\bfseries}}
\renewcommand\subsubsection{\@startsection{subsubsection}{3}{\z@}%
                                   {-2.3ex\@plus -1ex \@minus -.5ex}%
                                   {1ex \@plus .2ex \@minus .2ex}%
                                   {\normalfont\normalsize\bfseries}}
\renewcommand\paragraph{\@startsection{paragraph}{4}{\z@}%
                                   {1.75ex \@plus1ex \@minus.2ex}%
                                   {-1em}%
                                   {\normalfont\normalsize\bfseries}}
\renewcommand\subparagraph{\@startsection{subparagraph}{5}{\parindent}%
                                   {1.75ex \@plus1ex \@minus .2ex}%
                                   {-1em}%
                                   {\normalfont\normalsize\bfseries}}

\def\fnum@figure{\textbf{\figurename\nobreakspace\thefigure}}
\def\fnum@table{\textbf{\tablename\nobreakspace\thetable}}

\long\def\@makecaption#1#2{%
  \vskip\abovecaptionskip
  \sbox\@tempboxa{\small #1. #2}%
  \ifdim \wd\@tempboxa >\hsize
    \small #1. #2\par
  \else
    \global \@minipagefalse
    \hb@xt@\hsize{\hfil\box\@tempboxa\hfil}%
  \fi
  \vskip\belowcaptionskip}

\renewenvironment{thebibliography}[1]{%
\begin{oldthebibliography}{#1}%
\small%
\raggedright%
\setlength{\itemsep}{5pt plus 0.2ex minus 0.05ex}%
}%
{%
\end{oldthebibliography}%
}

\begin{document}


\title{\boldmath Thermo field dynamics approach for entropy of spacetime}

\author[a]{Yu-Zhu Chen,}
\author[a]{Wen-Du Li,}
\author[a,b,1]{and Wu-Sheng Dai}\note{daiwusheng@tju.edu.cn.}


\affiliation[a]{Department of Physics, Tianjin University, Tianjin 300072, P.R. China}
\affiliation[b]{LiuHui Center for Applied Mathematics, Nankai University \& Tianjin University, Tianjin 300072, P.R. China}









\abstract{A thermo field dynamics approach for calculating the entropy of a spacetime is
suggested. It is exemplified through the Rindler spacetime, the Milne
spacetime, the Boulware spacetime, and the Minkowski spacetime with a moving
mirror that the entropy of a spacetime is equal to the entropy of a thermo
quantum field with the same temperature of the spacetime we study. This in
fact suggests an thermo field dynamics approach of calculating the entropy of
a spacetime. In this approach, the entropy of a spacetime is an expectation
value of the entropy operator on a thermo vacuum state. The thermo vacuum
state is the vacuum state on the maximal manifold which is the maximal
analytic extension of the spacetime we study.}


\maketitle
\flushbottom


\section{Introduction}

Thermo field dynamics is a quantum field theory at finite temperature
\cite{takahashi1996thermo,hashizume2013understanding}. In thermo field
dynamics, the statistical average of a physical quantity is represented by an
expectation value of an  operator expressed by field operators on a
temperature-dependent thermo vacuum state $\left\vert 0\left(  \beta\right)
\right\rangle $. For example, the entropy is the expectation value of an
entropy operator on a thermo vacuum state $\left\vert 0\left(  \beta\right)
\right\rangle $. In this paper, we apply thermo field dynamics to calculate
the entropy of spacetime.

In our scheme, for the calculation of the entropy of a spacetime, the entropy
operator is constructed by thermo field dynamics and the thermo vacuum state
$\left\vert 0\left(  \beta\right)  \right\rangle $ is defined as the vacuum
state on the maximal manifold which is the maximal analytic extension of the
spacetime we study.

The validity of the approach is exemplified by applying the approach to
calculate the entropy of four kinds of spacetimes: the Rindler spacetime, the
Milne spacetime, the Boulware spacetime, and the Minkowski spacetime with a
moving mirror. We show that the result agrees well with literature.

Many approaches are developed for the calculation of the entropy of a
spacetime \cite{wald2002thermodynamics}. The first calculation is due to the
similarity between the property of a black hole and the law of thermodynamics
\cite{bardeen1973four,bekenstein1973black}. The entropy of a black hole is
then calculated by Euclidean quantum field theory \cite{gibbons1977action}. By
regarding the entropy of a black hole as an entanglement entropy between the
exterior and interior of the horizon, one can also calculate the entropy of a
spacetime \cite{callan1994geometric,holzhey1994geometric}. The black hole
entropy can also be calculated by quantum geometry \cite{ashtekar1998quantum}.
In string theory, the entropy of the $AdS$ spacetime is obtained through the
$AdS/CFT$ correspondence \cite{aharony2000large} due to the equivalence
between the $n+1$-dimensional $AdS$ and the conformal field theory on the
boundary \cite{maldacena1999large}. A relation between the black hole entropy
and the conformal field on the horizon is discussed in ref.
\cite{carlip2000black}. In ref. \cite{navarro2000ads}, the correspondence
between $AdS_{2}$ and $CFT_{1}$ is provided. The entropy of the $BTZ$ black
hole is discussed in ref. \cite{carlip1998we}. The entropy of a spherically
symmetric system is given by the Liouville theory in ref.
\cite{giacomini2003black}. The central charge of the conformal field which is
related to the entropy is obtained according to the symmetry of the system in
ref. \cite{silva2002black}. The entropy can also be obtained by counting
$D$-brane states \cite{strominger1996microscopic}. A derivation of the entropy
based on the microstates of $CFT$ is given by \cite{strominger1998black}. In
loop quantum gravity \cite{ashtekar1998quantum}, the\ entropy is obtained by
counting the microscopic states of a black hole and some efforts are devoted
to match the loop-quantum-gravity entropy with the Bekenstein-Hawking entropy
\cite{ghosh2011black,jacobson2007note,domagala2004black,rovelli1996black,agullo2009combinatorics,kloster2008phase}%
. The entropy of a black hole can also be calculated as a Noether charge
associated with the horizon Killing field
\cite{wald1993black,clunan2004gauss,vollick2007noether}.

In section \ref{review}, we provide a brief review of thermo field dynamics.
In section \ref{EoS}, we construct the thermo field dynamics approach for the
calculation of the entropy of a spacetime. In sections \ref{Rindler},
\ref{Milne}, \ref{Boulware}, and \ref{mirror}, as exemplification, we
calculate the entropy of the Rindler spacetime, the Milne spacetime, the
Boulware spacetime, and the Minkowski spacetime with a moving mirror. The
conclusion is given in section \ref{conclusion}.

\section{Calculating entropy: A brief review of thermo field dynamics
\label{review}}

In this section, we provide a brief review of thermo field dynamics
\cite{takahashi1996thermo,das1997finite,kapusta1993finite}.

In thermo field dynamics, the statistical average of a physical operator
$\hat{F}$ at a temperature $T=1/\beta$ is replaced by an expectation value of
$\hat{F}$ on a thermo vacuum state:%
\begin{equation}
\left\langle \hat{F}\right\rangle =\frac{1}{Z\left(  \beta\right)  }Tr\left(
\hat{F}e^{-\beta\hat{H}}\right)  =\left\langle 0\left(  \beta\right)
\right\vert \hat{F}\left\vert 0\left(  \beta\right)  \right\rangle ,
\label{1.1}%
\end{equation}
where the thermo vacuum state $\left\vert 0\left(  \beta\right)  \right\rangle
$ is a temperature-dependent vacuum state defined by
\begin{equation}
\left\vert 0\left(  \beta\right)  \right\rangle =\frac{1}{\sqrt{Z\left(
\beta\right)  }}\sum_{n}e^{-\beta E_{n}}\left\vert n\right\rangle \left\vert
\tilde{n}\right\rangle . \label{1.2}%
\end{equation}
Here, for zero chemical potential cases, $Z\left(  \beta\right)  =\sum
_{n}e^{-\beta E_{n}}$ is the partition function, the eigenvalue $E_{n}$ and
eigenvector $\left\vert n\right\rangle $ are determined by $\hat{H}\left\vert
n\right\rangle =E_{n}\left\vert n\right\rangle $, and $\left\vert \tilde
{n}\right\rangle $ is the eigenvector of the corresponding fictitious system
\cite{takahashi1996thermo}. The fictitious system represented by the tilde
state $\left\vert \tilde{n}\right\rangle $ is interpreted as the hole of the
physical particle in $\left\vert n\right\rangle $ \cite{takahashi1996thermo}.

To calculate the entropy of a spacetime by thermo field dynamics is just to
calculate the thermo vacuum expectation value of the entropy operator of the
spacetime, i.e.,%
\begin{equation}
S=\left\langle 0\left(  \beta\right)  \right\vert \hat{S}\left\vert 0\left(
\beta\right)  \right\rangle . \label{Sav}%
\end{equation}

The entropy operator in thermo field dynamics reads \cite{takahashi1996thermo}%
\begin{align}
\hat{S}  &  =-\sum\limits_{\mathbf{p}}\left[  b_{\mathbf{p}}^{\dag
}b_{\mathbf{p}}\ln\sinh^{2}\theta_{\mathbf{p}}\left(  \beta\right)
-b_{\mathbf{p}}b_{\mathbf{p}}^{\dag}\ln\cosh^{2}\theta_{\mathbf{p}}\left(
\beta\right)  \right]  ,\text{ \ \ bosonic,}\label{Sbose}\\
\hat{S}  &  =-\sum\limits_{\mathbf{p}}\left[  b_{\mathbf{p}}^{\dag
}b_{\mathbf{p}}\ln\sin^{2}\theta_{\mathbf{p}}\left(  \beta\right)
-b_{\mathbf{p}}b_{\mathbf{p}}^{\dag}\ln\cos^{2}\theta_{\mathbf{p}}\left(
\beta\right)  \right]  ,\text{\ \ \ \ \ fermionic,} \label{Sfermi}%
\end{align}
where $b_{\mathbf{p}}^{\dag}$ and $b_{\mathbf{p}}$ are creation and
annihilation operators and \cite{takahashi1996thermo}%
\begin{align}
\sinh^{2}\theta_{\mathbf{p}}\left(  \beta\right)   &  =\frac{1}{e^{\beta
\Omega_{\mathbf{p}}}-1},\text{\ \ \ bosonic,}\label{1.4}\\
\sin^{2}\theta_{\mathbf{p}}\left(  \beta\right)   &  =\frac{1}{e^{\beta
\Omega_{\mathbf{p}}}+1},\text{\ \ \ fermionic,}%
\end{align}
with $\Omega_{\mathbf{p}}=\left\vert \mathbf{p}\right\vert $, are
Bose-Einstein and Fermi-Dirac distributions, respectively.

\section{Thermo field dynamics approach for entropy of spacetime \label{EoS}}

In this section, we construct the approach for calculating the entropy in the
frame of thermo field dynamics. In next sections, as exemplification, we will
calculate the entropy of the Rindler spacetime, the Milne spacetime, the
Boulware spacetime, and the Minkowski spacetime with a moving mirror and
compare them with literature.

The key step in thermo field dynamics is to construct the thermo vacuum state
$\left\vert 0\left(  \beta\right)  \right\rangle $. To calculate the entropy
of a spacetime, we choose the thermo vacuum state as the vacuum state on the
maximal manifold, or, the maximal spacetime.

The maximal manifold is the maximal analytic extension of the spacetime we
study. In the maximal manifold every geodesic is either of infinite length in
both directions or else ends or begins on a singularity
\cite{ohanian2013gravitation}. The maximal manifold is described by the
complete atlas \cite{hawking1973large}. The spacetime we study is a
submanifold of the maximal manifold, and the maximal manifold can be achieved
by the maximal analytic extension of the spacetime we study. Take the Rindler
spacetime as an example. The Rindler spacetime is the region that an
accelerated observer with constant acceleration sees. The boundary of the
Rindler spacetime has no singularity at all. The Rindler spacetime can be
analytically continued to the Minkowski spacetime by a coordinates
transformation \cite{mukhanov2007introduction}. The Minkowski spacetime covers
not only the Rindler spacetime, but also the region besides the Rindler
spacetime, or, the Rindler spacetime is a subspace or submanifold of the
Minkowski spacetime. No other spacetime covers larger region than the
Minkowski spacetime does. Therefore, the Minkowski spacetime is a maximal
manifold. Moreover, the Minkowski spacetime is the unique maximal spacetime
corresponding to the Rindler spacetime according to the uniqueness of analytic continuation.

After analytically continuating the spacetime we study to the corresponding
maximal manifold, we have two kinds of spacetimes. On each of them, we can
define a set of creation and annihilation operators, respectively: on the
maximal manifold, $a_{\omega}^{\dag}$ and $a_{\omega}$; on the spacetime we
study, $b_{\Omega}^{\dag}$ and $b_{\Omega}$. Furthermore, we can define two
vacuum states by the corresponding annihilation operators:
\begin{equation}
a_{\omega}\left\vert 0_{a}\right\rangle =0
\end{equation}
defines the vacuum state $\left\vert 0_{a}\right\rangle $ on the maximal
manifold;
\begin{equation}
b_{\Omega}\left\vert 0_{b}\right\rangle =0
\end{equation}
defines the vacuum state $\left\vert 0_{b}\right\rangle $ on the spacetime we study.

In the approach, the thermo vacuum state is chosen as the vacuum state on the
maximal manifold, $\left\vert 0_{a}\right\rangle $.

Concretely, consider a real scalar field. Define creation and annihilation
operators, $b_{\Omega}^{\dag}$ and $b_{\Omega}$,$\ $on the spacetime we study,
such as the Rindler spacetime, etc. Furthermore, define creation and
annihilation operators, $a_{\omega}^{\dag}$ and $a_{\omega}$, on the
corresponding maximal manifold. Note that $\omega\in\left(  -\infty
,\infty\right)  $ and $\Omega\in\left(  -\infty,\infty\right)  $. For
conformally flat cases, these two sets of creation and annihilation operators
can be related by the Bogoliubov transformation
\cite{mukhanov2007introduction,mukhanov2004introduction},
\begin{align}
b_{\Omega}  &  =\int_{-\infty}^{\infty}d\omega\left(  \alpha_{\omega\Omega
}a_{\omega}-\beta_{\omega\Omega}a_{\omega}^{\dag}\right)  ,\nonumber\\
b_{\Omega}^{\dag}  &  =\int_{-\infty}^{\infty}d\omega\left(  \alpha
_{\omega\Omega}^{\ast}a_{\omega}^{\dag}-\beta_{\omega\Omega}^{\ast}a_{\omega
}\right)  . \label{2.0.0}%
\end{align}
The transformation coefficients $\alpha_{\omega\Omega}$ and $\beta
_{\omega\Omega}$ satisfy%
\begin{equation}
\int_{-\infty}^{\infty}d\omega\left(  \alpha_{\omega\Omega}\alpha
_{\omega\Omega^{\prime}}^{\ast}-\beta_{\omega\Omega}\beta_{\omega
\Omega^{\prime}}^{\ast}\right)  =\delta\left(  \Omega-\Omega^{\prime}\right)
. \label{2.0.1}%
\end{equation}
Particularly, for $\Omega=\Omega^{\prime}$, we have%
\begin{equation}
\int_{-\infty}^{\infty}d\omega\left(  \left\vert \alpha_{\omega\Omega
}\right\vert ^{2}-\left\vert \beta_{\omega\Omega}\right\vert ^{2}\right)
=\delta\left(  0\right)  . \label{2.0.2}%
\end{equation}

For a real scalar field, by eqs. (\ref{Sbose}) and (\ref{Sav}), the entropy
$S=\left\langle 0_{a}\right\vert \hat{S}\left\vert 0_{a}\right\rangle $, the
expectation value of the entropy operator $\hat{S}$ on the vacuum state
$\left\vert 0_{a}\right\rangle $:%
\begin{equation}
S=-\sum_{\mathbf{p}}\left[  \left\langle 0_{a}\right\vert b_{\Omega}^{\dag
}b_{\Omega}\left\vert 0_{a}\right\rangle \ln\sinh^{2}\theta_{\mathbf{p}%
}\left(  \beta\right)  -\left\langle 0_{a}\right\vert b_{\Omega}b_{\Omega
}^{\dag}\left\vert 0_{a}\right\rangle \ln\cosh^{2}\theta_{\mathbf{p}}\left(
\beta\right)  \right]  . \label{SaSa}%
\end{equation}

Substituting the Bogoliubov transformation (\ref{2.0.0}) into eq.
(\ref{SaSa}), we arrive at
\begin{equation}
S=\sum_{\mathbf{p}}\int_{-\infty}^{\infty}d\omega\left[  \left\vert
\alpha_{\omega\Omega}\right\vert ^{2}\ln\cosh^{2}\theta_{\mathbf{p}}\left(
\beta\right)  -\left\vert \beta_{\omega\Omega}\right\vert ^{2}\ln\sinh
^{2}\theta_{\mathbf{p}}\left(  \beta\right)  \right]  . \label{2.0.3}%
\end{equation}
Substituting the Bose-Einstein distribution (\ref{1.4}) and $\cosh^{2}%
\theta_{\mathbf{p}}\left(  \beta\right)  =e^{\beta\Omega_{\mathbf{p}}}/\left(
e^{\beta\Omega_{\mathbf{p}}}-1\right)  $ which is given by eq. (\ref{1.4})
into eq. (\ref{2.0.3}), we obtain
\begin{equation}
S=\sum_{\mathbf{p}}\left(  \ln\frac{e^{\beta\Omega_{\mathbf{p}}}}%
{e^{\beta\Omega_{\mathbf{p}}}-1}\int_{-\infty}^{\infty}d\omega\left\vert
\alpha_{\omega\Omega}\right\vert ^{2}-\ln\frac{1}{e^{\beta\Omega_{\mathbf{p}}%
}-1}\int_{-\infty}^{\infty}d\omega\left\vert \beta_{\omega\Omega}\right\vert
^{2}\right)  , \label{2.0.4}%
\end{equation}
where $\Omega_{\mathbf{p}}=\left\vert \Omega\right\vert $.

In a word, once the Bogoliubov transformation (\ref{2.0.0}) is constructed,
one can directly obtain the entropy by eq. (\ref{2.0.4}).

It should be emphasized that though the Bogoliubov transformation
(\ref{2.0.0}) is valid for conformally flat cases, the method constructed
above can be applied to more general cases. We will discuss the application of
this method to non-conformally flat cases further later.

\section{Entropy of Rindler Spacetime \label{Rindler}}

\textit{The Rindler spacetime}. The Rindler spacetime is the region that an
accelerated observer with constant acceleration $a$ sees, characterized by the
metric
\cite{mukhanov2007introduction,ohanian2013gravitation,crispino2008unruh}
\begin{equation}
ds^{2}=e^{2a\xi}\left(  d\tau^{2}-d\xi^{2}\right)  -dy^{2}-dz^{2}.
\label{2.0.5}%
\end{equation}

The maximal manifold of the Rindler spacetime is the Minkowski spacetime:
\begin{equation}
ds^{2}=dt^{2}-dx^{2}-dy^{2}-dz^{2}. \label{2.0.6}%
\end{equation}
The coordinates in the metric (\ref{2.0.5}), $\left(  \tau,\xi\right)  $, and
the coordinates in the metric (\ref{2.0.6}), $\left(  t,x\right)  $, are
connected by \cite{mukhanov2007introduction}%
\begin{equation}
t=\frac{1}{a}e^{a\xi}\sinh\left(  a\tau\right)  ,\text{ \ }x=\frac{1}%
{a}e^{a\xi}\cosh\left(  a\tau\right)  .
\end{equation}

\textit{The real scalar field}. Consider real scaler fields in the Minkowski
spacetime and in the Rindler spacetime. In the Minkowski spacetime, denote the
creation and annihilation operators as $a_{\omega}^{\dag}$ and $a_{\omega}$;
in the Rindler spacetime, denote the creation and annihilation operators as
$b_{\Omega}^{\dag}$ and $b_{\Omega}$.

The vacuum state in the Minkowski spacetime $\left\vert 0_{M}\right\rangle $
is defined by%
\begin{equation}
a_{\omega}\left\vert 0_{M}\right\rangle =0.
\end{equation}
For massless cases, we have $\left\vert \omega\right\vert =\left\vert
\mathbf{k}\right\vert $ in the Minkowski spacetime and $\left\vert
\Omega\right\vert =\left\vert \mathbf{p}\right\vert $ in the Rindler spacetime.

\textit{The entropy of the Rindler spacetime}.\textit{ }As analyzed in section
\ref{EoS}, the entropy $S_{R}=\left\langle 0_{M}\right\vert \hat{S}%
_{R}\left\vert 0_{M}\right\rangle $ is the expectation value of $\hat{S}_{R}$
on the corresponding thermo vacuum --- the vacuum of the Minkowski spacetime
$\left\vert 0_{M}\right\rangle $.

The Bogoliubov transformation coefficients in eq. (\ref{2.0.0})\textbf{,}
$\alpha_{\omega\Omega}$ and $\beta_{\omega\Omega}$, satisfy
\cite{mukhanov2007introduction}
\begin{equation}
\left\vert \alpha_{\omega\Omega}\right\vert ^{2}=e^{\beta\left\vert
\Omega\right\vert }\left\vert \beta_{\omega\Omega}\right\vert ^{2},
\label{2.3}%
\end{equation}
where $\beta=2\pi/a$. Substituting eq. (\ref{2.3}) into eq. (\ref{2.0.2}), we
arrive at%
\begin{equation}
\int_{-\infty}^{\infty}d\omega\left\vert \beta_{\omega\Omega}\right\vert
^{2}=\frac{1}{e^{\beta\left\vert \Omega\right\vert }-1}\delta\left(  0\right)
. \label{2.3.2}%
\end{equation}
Substituting the relation (\ref{2.3}) with $\Omega_{\mathbf{p}}=\left\vert
\Omega\right\vert $ into eq. (\ref{2.0.4}) and using (\ref{2.3.2}), we obtain%
\begin{equation}
S_{R}=\delta\left(  0\right)  \sum_{\mathbf{p}}\left[  \beta\left\vert
\Omega\right\vert \frac{e^{\beta\left\vert \Omega\right\vert }}{e^{\beta
\left\vert \Omega\right\vert }-1}-\ln\left(  e^{\beta\left\vert \Omega
\right\vert }-1\right)  \right]  . \label{2.6}%
\end{equation}

For $1+3$ dimensions, converting the summation into an integral,
$\sum_{\mathbf{p}}\rightarrow4\pi\int_{0}^{\infty}\Omega^{2}d\Omega$, where
$\left\vert \Omega\right\vert =\left\vert \mathbf{p}\right\vert $, gives
\begin{equation}
S_{R}=\frac{16\pi^{5}}{45\beta^{3}}\delta\left(  0\right)  .
\end{equation}
Here $\delta\left(  0\right)  $ represents the volume of the Rindler
spacetime: $\delta\left(  0\right)  =\left.  \int\frac{d^{3}x}{\left(
2\pi\right)  ^{3}}e^{i\mathbf{p\cdot x}}\right\vert _{\mathbf{p}=\mathbf{0}%
}=\frac{1}{\left(  2\pi\right)  ^{3}}\int d^{3}x=\frac{1}{\left(  2\pi\right)
^{3}}V$.

The entropy then reads%
\begin{equation}
S_{R}=\frac{2\pi^{2}}{45}VT^{3}. \label{2.8}%
\end{equation}
This result agrees with the result given in ref.
\cite{susskind2005introduction}.

\section{Entropy of Milne spacetime \label{Milne}}

\textit{The }$1+1$-dimensional \textit{Milne spacetime}. The Milne spacetime
is the region of the future lightcone of a given point in the Minkowski
spacetime \cite{emelyanov2014local,emelyanov2014freely}, which can be, for
example, regarded as the empty universe of the FLRW model
\cite{carroll2004spacetime}. The Milne spacetime is characterized by the
metric
\begin{equation}
ds^{2}=e^{2a\tau}\left(  d\tau^{2}-d\xi^{2}\right)  . \label{E.1}%
\end{equation}

The maximal manifold of the Milne spacetime is the Minkowski spacetime
\begin{equation}
ds^{2}=dt^{2}-dx^{2}. \label{E.2}%
\end{equation}
The coordinates in the metric (\ref{E.1}), $\left(  \tau,\xi\right)  $, and
the coordinates in the metric (\ref{E.2}), $\left(  t,x\right)  $, are
connected by
\begin{equation}
t=\frac{1}{a}e^{a\tau}\cosh\left(  a\xi\right)  ,\text{ \ }x=\frac{1}%
{a}e^{a\tau}\sinh\left(  a\xi\right)  .
\end{equation}

\textit{The real scalar field}. The creation and annihilation operators of a
real scalar field in the Minkowski spacetime are $a_{\omega}^{\dag}$ and
$a_{\omega}$, and in the Milne spacetime are $b_{\Omega}^{\dag}$ and
$b_{\Omega}$.

The vacuum state in the Minkowski spacetime $\left\vert 0_{M}\right\rangle $
is defined by%
\begin{equation}
a_{\omega}\left\vert 0_{M}\right\rangle =0.
\end{equation}

\textit{The entropy of the Milne spacetime}.\textit{ }Similarly, the entropy
$S_{Mil}$ $=\left\langle 0_{M}\right\vert \hat{S}_{Mil}\left\vert
0_{M}\right\rangle $ is the expectation value of $\hat{S}_{Mil}$ on the
corresponding thermo vacuum --- the vacuum of the Minkowski spacetime
$\left\vert 0_{M}\right\rangle $.

The Bogoliubov transformation coefficients in eq. (\ref{2.0.0})\textbf{,}
$\alpha_{\omega\Omega}$ and $\beta_{\omega\Omega}$, satisfy
\begin{equation}
\left\vert \alpha_{\omega\Omega}\right\vert ^{2}=e^{\beta\left\vert
\Omega\right\vert }\left\vert \beta_{\omega\Omega}\right\vert ^{2},
\label{E.4}%
\end{equation}
where $\beta=2\pi/a$. By the same procedure in section \ref{Rindler}, we
obtain the entropy:%
\begin{equation}
S_{Mil}=\sum_{\mathbf{p}}\left(  \ln\frac{e^{\beta\left\vert \Omega\right\vert
}}{e^{\beta\left\vert \Omega\right\vert }-1}e^{\beta\left\vert \Omega
\right\vert }\int_{-\infty}^{\infty}d\omega\left\vert \beta_{\omega\Omega
}\right\vert ^{2}-\ln\frac{1}{e^{\beta\left\vert \Omega\right\vert }-1}%
\int_{-\infty}^{\infty}d\omega\left\vert \beta_{\omega\Omega}\right\vert
^{2}\right)  . \label{E.6}%
\end{equation}
We then have%
\begin{equation}
S_{Mil}=\delta\left(  0\right)  \sum_{\mathbf{p}}\left[  \beta\left\vert
\Omega\right\vert \frac{e^{\beta\left\vert \Omega\right\vert }}{e^{\beta
\left\vert \Omega\right\vert }-1}-\ln\left(  e^{\beta\left\vert \Omega
\right\vert }-1\right)  \right]  . \label{E.7}%
\end{equation}

Converting the summation into an integral $\sum_{\mathbf{p}}$ $\rightarrow
\int_{0}^{\infty}d\Omega$, where $\Omega=\left\vert \mathbf{p}\right\vert $,
give
\begin{equation}
S_{Mil}=\frac{\pi^{2}}{3\beta}\delta\left(  0\right)  .
\end{equation}
Here $\delta\left(  0\right)  $ represents the volume of the Milne spacetime:
$\delta\left(  0\right)  =L/\left(  2\pi\right)  $.

The entropy then reads%
\begin{equation}
S_{Mil}=\frac{\pi}{6\beta}L.
\end{equation}

\section{Entropy of Boulware spacetime \label{Boulware}}

\textit{The Boulware spacetime.} The Boulware spacetime is the region outside
a Schwarzschild black hole which is described by the tortoise coordinate
$\left(  t,r_{\ast}\right)  $ \cite{mukhanov2007introduction},%
\begin{equation}
ds^{2}=\left(  1-\frac{2M}{r}\right)  \left(  dt^{2}-dr_{\ast}^{2}\right)
-r^{2}\left(  d^{2}\theta+\sin^{2}\theta d\phi^{2}\right)  , \label{2.10}%
\end{equation}
where $r_{\ast}=r+2M\ln\frac{r-2M}{2M}$.

The maximal manifold of the Boulware spacetime is the Kruskal-Szekeres
spacetime described by the Kruskal-Szekeres coordinate $\left(  \tilde
{t},\tilde{r}\right)  $,%
\begin{equation}
ds^{2}=\frac{2M}{r}e^{-r/\left(  2M\right)  }\left(  d\tilde{t}^{2}-d\tilde
{r}^{2}\right)  -r^{2}\left(  d^{2}\theta+\sin^{2}\theta d\phi^{2}\right)  .
\label{2.11}%
\end{equation}
The Kruskal-Szekeres coordinate $\left(  \tilde{t},\tilde{r}\right)  $ and the
tortoise coordinate $\left(  t,r_{\ast}\right)  $ are connected by
\cite{mukhanov2007introduction,ohanian2013gravitation}%
\begin{equation}
\tilde{t}=4Me^{r_{\ast}/\left(  4M\right)  }\sinh\frac{t}{4M},\text{\ }%
\tilde{r}=4Me^{r_{\ast}/\left(  4M\right)  }\cosh\frac{t}{4M}. \label{2.12}%
\end{equation}

\textit{The real scalar field}. In the Kruskal-Szekeres spacetime, denote the
creation and annihilation operators as $a_{\omega}^{\dag}$ and $a_{\omega}$;
in the Boulware spacetime, denote the creation and annihilation operators as
$b_{\Omega}^{\dag}$ and $b_{\Omega}$.

The vacuum state in the Kruskal-Szekeres spacetime $\left\vert 0_{K}%
\right\rangle $ is defined by%
\begin{equation}
a_{\omega}\left\vert 0_{K}\right\rangle =0.
\end{equation}
For massless cases, we have $\left\vert \omega\right\vert =\left\vert
\mathbf{k}\right\vert $ in the Kruskal-Szekeres spacetime and $\left\vert
\Omega\right\vert =\left\vert \mathbf{p}\right\vert $ in the Boulware spacetime.

\textit{The entropy of the Boulware spacetime}.\textit{ }As above, the entropy
$S_{B}$ $=\left\langle 0_{K}\right\vert \hat{S}_{B}\left\vert 0_{K}%
\right\rangle $ is the expectation value of $\hat{S}_{B}$ on the corresponding
thermo vacuum --- the vacuum of the Kruskal-Szekeres spacetime $\left\vert
0_{K}\right\rangle $.

The coefficients, $\alpha_{\omega\Omega}$ and $\beta_{\omega\Omega}$,\ of the
Bogoliubov transformation between $a_{\omega}^{\dag}$, $a_{\omega}\ $and
$b_{\Omega}^{\dag}$, $b_{\Omega}$ satisfy
\cite{mukhanov2007introduction,hawking1975particle}
\begin{equation}
\left\vert \alpha_{\omega\Omega}\right\vert ^{2}=e^{\beta\left\vert
\Omega\right\vert }\left\vert \beta_{\omega\Omega}\right\vert ^{2},
\label{2.13}%
\end{equation}
where $\beta=8\pi M$. By the same procedure in section \ref{Rindler}, we
obtain the entropy of the Boulware spacetime:%
\begin{equation}
S_{B}=\frac{16\pi^{5}}{45\beta^{3}}\delta\left(  0\right)  =\frac{2\pi^{2}%
}{45\beta^{3}}V. \label{2.18}%
\end{equation}
Taking the volume as $V=\frac{4}{3}\pi r^{3}$ gives%
\begin{equation}
S_{B}=\frac{8\pi^{3}}{135\beta^{3}}r^{3}. \label{2.19}%
\end{equation}
This result agrees with the brick wall model \cite{t1985quantum}. In ref.
\cite{t1985quantum}, for calculating the entropy of a Schwarzschild black
hole, t' Hooft constructs the brick wall model. In the brick wall model, one
equates the entropy of the Bose gas outside the black hole with the entropy of
the Schwarzschild black hole.

\section{Entropy of Minkowski spacetime with a moving mirror \label{mirror}}

Placing a moving mirror in a spacetime is equivalent to setting an appropriate
boundary condition \cite{davies1977radiation,fulling1976radiation}.

\textit{The} $1+1$-dimensional \textit{Minkowski spacetime with a moving
mirror. }A\textit{ }Minkowski spacetime with a moving mirror as a boundary can
be characterized by the metric \cite{davies1977radiation,fulling1976radiation}%
\[
ds^{2}=dt^{2}-dx^{2},
\]
with the boundary
\begin{equation}
x=Z\left(  t\right)  =\left\{
\begin{array}
[c]{c}%
-t-Ae^{-2\kappa t}+B,\text{ \ }t>0\\
0,\text{ \ \ \ \ \ \ \ \ \ \ \ \ \ }t<0
\end{array}
\right.  , \label{2.20}%
\end{equation}
where $A$, $B$, and $\kappa$ are constants.

The corresponding maximal manifold of such a spacetime is the Minkowski
spacetime without boundaries. In the following, we call the Minkowski
spacetime without boundaries the \textit{in} spacetime and the Minkowski
spacetime with a moving mirror (\ref{2.20}) the \textit{out}\ spacetime. More
detailed descriptions can be found in refs
\cite{davies1977radiation,fulling1976radiation}.

\textit{The real scalar field}. In the \textit{in} spacetime (without
boundaries), denote the creation and annihilation operators as $a_{\omega
}^{\dag}$ and $a_{\omega}$; in the \textit{out} spacetime (with boundaries),
denote the creation and annihilation operators as $b_{\Omega}^{\dag}$ and
$b_{\Omega}$.

The vacuum state in the \textit{in} spacetime $\left\vert 0_{in}\right\rangle
$ is defined by%
\begin{equation}
a_{\omega}\left\vert 0_{in}\right\rangle =0.
\end{equation}

\textit{The entropy of the out spacetime}. The entropy of the \textit{out}
spacetime $S_{out}=\left\langle 0_{in}\right\vert \hat{S}_{out}\left\vert
0_{in}\right\rangle $ is the expectation value of $\hat{S}_{out}$ on the
corresponding thermo vacuum ------ the vacuum of the \textit{in} spacetime
$\left\vert 0_{in}\right\rangle $.

The coefficients, $\alpha_{\omega\Omega}$ and $\beta_{\omega\Omega}$, in the
Bogoliubov transformation between $a_{\omega}^{\dag}$, $a_{\omega}\ $and
$b_{\Omega}^{\dag}$, $b_{\Omega}$ satisfy \cite{fulling1976radiation}
\begin{equation}
\left\vert \alpha_{\omega\Omega}\right\vert ^{2}=e^{\beta\Omega}\left\vert
\beta_{\omega\Omega}\right\vert ^{2}, \label{43}%
\end{equation}
where $\beta\equiv2\pi/\kappa$. By the same procedure in the above sections,
we obtain the entropy:%

\begin{equation}
S_{out}=\sum_{p}\left[  \beta\Omega\frac{e^{\beta\Omega}}{e^{\beta\Omega}%
-1}-\ln\left(  e^{\beta\Omega}-1\right)  \right]  \delta\left(  0\right)  .
\label{46}%
\end{equation}
Converting the summation into an integral $\sum_{p}\rightarrow\int_{0}%
^{\infty}d\Omega$ with $\Omega=\left\vert p\right\vert $ gives
\begin{equation}
S_{out}=\frac{\pi^{2}}{3\beta}\delta\left(  0\right)  =\frac{\pi}{6\beta}L,
\label{48}%
\end{equation}
where $L$ is the volume.

In the \textit{out} spacetime, the observer sees particles created from the
mirror. The physical picture is that the particle in the Minkowski spacetime
(\textit{in} spacetime) moves to the mirror and then is reflected into the
\textit{out} spacetime. In other words, the \textit{out} spacetime is not an
isolated system, it can exchange particles and energy with the maximal
manifold. The moving mirror here plays a role similar to the horizon of a
black hole.

\section{Conclusion \label{conclusion}}

In this paper, we suggest a thermo-field-dynamics approach for the calculation
of the entropy of spacetimes. The approach is exemplified through calculating
the entropy of the Rindler spacetime, the Milne spacetime, the Boulware
spacetime, and the Minkowski spacetime with a moving mirror.

In this scheme, the entropy of a spacetime is an expectation value of the
entropy operator on a thermo vacuum state $\left\vert 0\left(  \beta\right)
\right\rangle $ which is the vacuum state of a quantum field in the maximal
manifold of the spacetime we study. Concretely, first analytically continue
the spacetime to its maximal manifold. Then define the field, in our case a
scalar field, on both the spacetime and its maximal manifold. The field at the
same point on the spacetime and on its maximal manifold, clearly, must be
equal to each other. This determines the coefficients of the Bogoliubov
transformation which connects the fields on the spacetime and its maximal manifold.

It is known that there is no generally accepted definition of the entropy of
spacetime. Historically, the entropy of spacetime was introduced by a
non-rigorous analogy between the horizon area and the entropy
\cite{bekenstein1994we,christodoulou1970reversible,penrose1971extraction,hawking1971gravitational}%
. Bekenstein showed that there is a way to understand the entropy of spacetime
just as that in statistical mechanics: the logarithm of the number of all the
possible states which can form a black hole. Moreover, there are many attempts
on the explanation and calculation of the entropy of spacetime. For example,
regard the entropy of a black hole as a Noether charge
\cite{wald1993black,clunan2004gauss,vollick2007noether}, explain the entropy
of a black hole as a topological contribution to the Euclidean action
integrals \cite{gibbons1977action,hawking1996nature}, explain the entropy as
the logarithm of the number of the states of the quantum states of the
geometry of the horizon in quantum geometry theory \cite{ashtekar1998quantum},
equal the entropy of a black hole with the logarithm of the number of the
states of the conformal field on the boundary of the black hole by the
$AdS/CFT$ correspondence
\cite{carlip2000black,aharony2000large,navarro2000ads}, or regard the entropy
as the logarithm of the number of\textbf{ }the $D$-brane states
\cite{strominger1996microscopic}, etc. An important explanation of the entropy
of spacetime is the entanglement entropy. The discovery of black-hole
radiation \cite{hawking1975particle} allows us to explain the entropy of
spacetime as an entanglement entropy \cite{bombelli1986quantum}. The quantum
field on the exterior of the black hole is in a mixed state. The entanglement
entropy comes from the quantum field correlations between the exterior and
interior of the horizon \cite{wald2002thermodynamics}. The entanglement
entropy has been adopted by many authors
\cite{callan1994geometric,holzhey1994geometric,susskind1994black}. The entropy
we have calculated above is the entropy of a thermal equilibrium massless Bose
gas outside the black hole. Such an entropy is essentially an entanglement entropy.

Moreover, if regarding the entropy of a spacetime as an entanglement entropy,
one encounters a problem --- the species problem \cite{bekenstein1994we}. The
Hawking radiation emits all species of particles and then the entropy should
depend on the number of the species of particles. There are some discussions
of the species problem. For example, Sorkin and `t Hooft suggest that the
entropy is a summation of all contributions of all species of particles
\cite{bekenstein1994we} and Jacobson \cite{jacobson1994black} and Susskind
\cite{susskind1994black} consider the possibility\ of correcting the
gravitational constant $G$. In the future research, to study the species
problem, we will consider the fermionic contribution to the entropy by the
thermo field dynamics approach.

The non-conformally flat case is also an important issue
\cite{emelyanov2014non,emelyanov2014local}. The method established in the
present paper can also be applied to such cases, which will be discussed
further later.

Furthermore, the thermo field dynamics approach suggested in the present paper
can also be applied to the calculation of other thermodynamic quantities of a
spacetime. All thermodynamic quantities are expectation values of their
corresponding operators which is defined in thermo field dynamics on the
thermo vacuum state $\left\vert 0\left(  \beta\right)  \right\rangle $. In
particular, the partition function $Z\left(  \beta\right)  =\left\langle
0\left(  \beta\right)  \right\vert e^{-\beta\hat{H}}\left\vert 0\left(
\beta\right)  \right\rangle $ is the Euclidean action of the system. This
means that we can study quantum field theory in curved space using the
approach. Moreover, the partition function is also the global heat kernel by
which one can calculate many quantities in quantum field theory, such as
one-loop effective actions, vacuum energies, and spectral counting functions
\cite{dai2009number,dai2010approach}.


\acknowledgments

We are very indebted to Dr G. Zeitrauman for his encouragement. We would like
to express our appreciation to Dr. Slava Emelyanov for his helpful comments
and suggestions. This work is supported in part by NSF of China under Grant
No. 11575125 and No. 11375128.










\providecommand{\href}[2]{#2}\begingroup\raggedright\endgroup


\begin{thebibliography}{10}

\bibitem{takahashi1996thermo}
Y.~Takahashi and H.~Umezawa, {\it Thermo field dynamics},  {\em International
  Journal of Modern Physics B} {\bf 10} (1996), no.~13n14 1755--1805.

\bibitem{hashizume2013understanding}
Y.~Hashizume and M.~Suzuki, {\it Understanding quantum entanglement by thermo
  field dynamics},  {\em Physica A: Statistical Mechanics and its Applications}
  {\bf 392} (2013), no.~17 3518--3530.

\bibitem{wald2002thermodynamics}
R.~M. Wald, {\it The thermodynamics of black holes},  in {\em Advances in the
  Interplay Between Quantum and Gravity Physics}, pp.~477--522.
\newblock Springer, 2002.

\bibitem{bardeen1973four}
J.~M. Bardeen, B.~Carter, and S.~W. Hawking, {\it The four laws of black hole
  mechanics},  {\em Communications in Mathematical Physics} {\bf 31} (1973),
  no.~2 161--170.

\bibitem{bekenstein1973black}
J.~D. Bekenstein, {\it Black holes and entropy},  {\em Physical Review D} {\bf
  7} (1973), no.~8 2333.

\bibitem{gibbons1977action}
G.~W. Gibbons and S.~W. Hawking, {\it Action integrals and partition functions
  in quantum gravity},  {\em Physical Review D} {\bf 15} (1977), no.~10 2752.

\bibitem{callan1994geometric}
C.~Callan and F.~Wilczek, {\it On geometric entropy},  {\em Physics Letters B}
  {\bf 333} (1994), no.~1 55--61.

\bibitem{holzhey1994geometric}
C.~Holzhey, F.~Larsen, and F.~Wilczek, {\it Geometric and renormalized entropy
  in conformal field theory},  {\em Nuclear Physics B} {\bf 424} (1994), no.~3
  443--467.

\bibitem{ashtekar1998quantum}
A.~Ashtekar, J.~Baez, A.~Corichi, and K.~Krasnov, {\it Quantum geometry and
  black hole entropy},  {\em Physical Review Letters} {\bf 80} (1998), no.~5
  904.

\bibitem{aharony2000large}
O.~Aharony, S.~S. Gubser, J.~Maldacena, H.~Ooguri, and Y.~Oz, {\it Large n
  field theories, string theory and gravity},  {\em Physics Reports} {\bf 323}
  (2000), no.~3 183--386.

\bibitem{maldacena1999large}
J.~Maldacena, {\it The large-n limit of superconformal field theories and
  supergravity},  {\em International journal of theoretical physics} {\bf 38}
  (1999), no.~4 1113--1133.

\bibitem{carlip2000black}
S.~Carlip, {\it Black hole entropy from horizon conformal field theory},  {\em
  Nuclear Physics B-Proceedings Supplements} {\bf 88} (2000), no.~1 10--16.

\bibitem{navarro2000ads}
J.~Navarro-Salas and P.~Navarro, {\it Ads 2/cft 1 correspondence and
  near-extremal black hole entropy},  {\em Nuclear Physics B} {\bf 579} (2000),
  no.~1 250--266.

\bibitem{carlip1998we}
S.~Carlip, {\it What we don't know about btz black hole entropy},  {\em
  Classical and Quantum Gravity} {\bf 15} (1998), no.~11 3609.

\bibitem{giacomini2003black}
A.~Giacomini and N.~Pinamonti, {\it Black hole entropy from classical liouville
  theory},  {\em Journal of High Energy Physics} {\bf 2003} (2003), no.~02 014.

\bibitem{silva2002black}
S.~Silva, {\it Black-hole entropy and thermodynamics from symmetries},  {\em
  Classical and Quantum Gravity} {\bf 19} (2002), no.~15 3947.

\bibitem{strominger1996microscopic}
A.~Strominger and C.~Vafa, {\it Microscopic origin of the bekenstein-hawking
  entropy},  {\em Physics Letters B} {\bf 379} (1996), no.~1 99--104.

\bibitem{strominger1998black}
A.~Strominger, {\it Black hole entropy from near-horizon microstates},  {\em
  Journal of High Energy Physics} {\bf 1998} (1998), no.~02 009.

\bibitem{ghosh2011black}
A.~Ghosh and A.~Perez, {\it Black hole entropy and isolated horizons
  thermodynamics},  {\em Physical review letters} {\bf 107} (2011), no.~24
  241301.

\bibitem{jacobson2007note}
T.~Jacobson, {\it A note on renormalization and black hole entropy in loop
  quantum gravity},  {\em Classical and Quantum Gravity} {\bf 24} (2007),
  no.~18 4875.

\bibitem{domagala2004black}
M.~Domagala and J.~Lewandowski, {\it Black-hole entropy from quantum geometry},
   {\em Classical and Quantum Gravity} {\bf 21} (2004), no.~22 5233.

\bibitem{rovelli1996black}
C.~Rovelli, {\it Black hole entropy from loop quantum gravity},  {\em Physical
  Review Letters} {\bf 77} (1996), no.~16 3288.

\bibitem{agullo2009combinatorics}
I.~Agull{\'o}, E.~F. Borja, J.~D{\'\i}az-Polo, E.~J. Villase{\~n}or, et~al.,
  {\it Combinatorics of the su (2) black hole entropy in loop quantum gravity},
   {\em Physical Review D} {\bf 80} (2009), no.~8 084006.

\bibitem{kloster2008phase}
S.~Kloster, J.~Brannlund, and A.~DeBenedictis, {\it Phase space and black-hole
  entropy of higher genus horizons in loop quantum gravity},  {\em Classical
  and Quantum Gravity} {\bf 25} (2008), no.~6 065008.

\bibitem{wald1993black}
R.~M. Wald, {\it Black hole entropy is the noether charge},  {\em Physical
  Review D} {\bf 48} (1993), no.~8 R3427.

\bibitem{clunan2004gauss}
T.~Clunan, S.~F. Ross, and D.~J. Smith, {\it On gauss--bonnet black hole
  entropy},  {\em Classical and Quantum Gravity} {\bf 21} (2004), no.~14 3447.

\bibitem{vollick2007noether}
D.~N. Vollick, {\it Noether charge and black hole entropy in modified theories
  of gravity},  {\em Physical Review D} {\bf 76} (2007), no.~12 124001.

\bibitem{das1997finite}
A.~Das, {\em Finite temperature field theory}, vol.~16.
\newblock World Scientific, 1997.

\bibitem{kapusta1993finite}
J.~I. Kapusta, {\em Finite-temperature field theory}.
\newblock Cambridge University Press, 1993.

\bibitem{ohanian2013gravitation}
H.~C. Ohanian and R.~Ruffini, {\em Gravitation and spacetime}.
\newblock Cambridge University Press, 2013.

\bibitem{hawking1973large}
S.~W. Hawking, {\em The large scale structure of space-time}, vol.~1.
\newblock Cambridge university press, 1973.

\bibitem{mukhanov2007introduction}
V.~Mukhanov and S.~Winitzki, {\em Introduction to quantum effects in gravity}.
\newblock Cambridge University Press, 2007.

\bibitem{mukhanov2004introduction}
V.~Mukhanov and S.~Winitzki, ``\textsl{Introduction to Quantum Fields in
  Classical Backgrounds}.'' This is a draft version of
  \cite{mukhanov2007introduction}, 2004.

\bibitem{crispino2008unruh}
L.~C. Crispino, A.~Higuchi, and G.~E. Matsas, {\it The unruh effect and its
  applications},  {\em Reviews of Modern Physics} {\bf 80} (2008), no.~3 787.

\bibitem{susskind2005introduction}
L.~Susskind and J.~Lindesay, {\em An introduction to black holes, information
  and the string theory revolution}.
\newblock World Scientific, 2005.

\bibitem{emelyanov2014local}
S.~Emelyanov, {\it Local thermal observables in spatially open frw spaces},
  {\em arXiv preprint arXiv:1406.3360} (2014).

\bibitem{emelyanov2014freely}
S.~Emelyanov, {\it Freely moving observer in (quasi) anti--de sitter space},
  {\em Physical Review D} {\bf 90} (2014), no.~4 044039.

\bibitem{carroll2004spacetime}
S.~M. Carroll, {\em Spacetime and geometry. An introduction to general
  relativity}, vol.~1.
\newblock 2004.

\bibitem{hawking1975particle}
S.~W. Hawking, {\it Particle creation by black holes},  {\em Communications in
  mathematical physics} {\bf 43} (1975), no.~3 199--220.

\bibitem{t1985quantum}
G.~t~Hooft, {\it On the quantum structure of a black hole},  {\em Nuclear
  Physics B} {\bf 256} (1985) 727--745.

\bibitem{davies1977radiation}
P.~Davies and S.~Fulling, {\it Radiation from moving mirrors and from black
  holes},  {\em Proceedings of the Royal Society of London. A. Mathematical and
  Physical Sciences} {\bf 356} (1977), no.~1685 237--257.

\bibitem{fulling1976radiation}
S.~Fulling and P.~Davies, {\it Radiation from a moving mirror in two
  dimensional space-time: conformal anomaly},  {\em Proceedings of the Royal
  Society of London. A. Mathematical and Physical Sciences} {\bf 348} (1976),
  no.~1654 393--414.

\bibitem{bekenstein1994we}
J.~D. Bekenstein, {\it Do we understand black hole entropy?},  {\em arXiv
  preprint gr-qc/9409015} (1994).

\bibitem{christodoulou1970reversible}
D.~Christodoulou, {\it Reversible and irreversible transformations in
  black-hole physics},  {\em Physical Review Letters} {\bf 25} (1970), no.~22
  1596.

\bibitem{penrose1971extraction}
R.~Penrose and R.~Floyd, {\it Extraction of rotational energy from a black
  hole},  {\em Nature} {\bf 229} (1971), no.~6 177--179.

\bibitem{hawking1971gravitational}
S.~W. Hawking, {\it Gravitational radiation from colliding black holes},  {\em
  Physical Review Letters} {\bf 26} (1971) 1344--1346.

\bibitem{hawking1996nature}
S.~W. Hawking, R.~Penrose, and M.~Atiyah, {\em The nature of space and time}.
\newblock Princeton University Press Princeton, 1996.

\bibitem{bombelli1986quantum}
L.~Bombelli, R.~K. Koul, J.~Lee, and R.~D. Sorkin, {\it Quantum source of
  entropy for black holes},  {\em Physical Review D} {\bf 34} (1986), no.~2
  373.

\bibitem{susskind1994black}
L.~Susskind and J.~Uglum, {\it Black hole entropy in canonical quantum gravity
  and superstring theory},  {\em Physical Review D} {\bf 50} (1994), no.~4
  2700.

\bibitem{jacobson1994black}
T.~Jacobson, {\it Black hole entropy and induced gravity},  {\em arXiv preprint
  gr-qc/9404039} (1994).

\bibitem{emelyanov2014non}
S.~Emelyanov, {\it Non-unitarity or hidden observables?},  {\em arXiv preprint
  arXiv:1410.6149} (2014).

\bibitem{dai2009number}
W.-S. Dai and M.~Xie, {\it The number of eigenstates: counting function and
  heat kernel},  {\em Journal of High Energy Physics} {\bf 2009} (2009), no.~02
  033.

\bibitem{dai2010approach}
W.-S. Dai and M.~Xie, {\it An approach for the calculation of one-loop
  effective actions, vacuum energies, and spectral counting functions},  {\em
  Journal of High Energy Physics} {\bf 2010} (2010), no.~6 1--29.



\end{thebibliography}

\end{document}